\documentclass[fleqn,usenatbib]{mnras}

\usepackage{newtxtext,newtxmath}

\usepackage[T1]{fontenc}

\DeclareRobustCommand{\VAN}[3]{#2}
\let\VANthebibliography\thebibliography
\def\thebibliography{\DeclareRobustCommand{\VAN}[3]{##3}\VANthebibliography}

\usepackage{graphicx}	
\usepackage{amsmath}	

\title[\emph{Suzaku} spectrum of 2S 0921-63]{Broad-band spectral analysis of LMXB 2S 0921-63 with \emph{Suzaku}}

\author[P. Sharma, C. Jain, and A. Dutta.]{
Prince Sharma,$^{1}$\thanks{E-mail: princerajsharma31@gmail.com}
Chetana Jain$^{2}$\thanks{E-mail: chetanajain11@gmail.com}
and Anjan Dutta$^{1}$\thanks{E-mail: dutta.anjan33@gmail.com}
\\
$^{1}$Department of Physics and Astrophysics, University of Delhi, Delhi 110007, India\\
$^{2}$Hansraj College, University of Delhi, Delhi 110007, India\\
}

\date{Accepted XXX. Received YYY; in original form ZZZ}

\pubyear{2022}

\begin{document}
\label{firstpage}
\pagerange{\pageref{firstpage}--\pageref{lastpage}}
\maketitle

\begin{abstract}
We present the broad-band spectral analysis of the low-mass X-ray binary 2S 0921-63 by using the \emph{Suzaku} archival data covering the orbital phase between 0.31 and 1.16 during four close observations. It is the first time that a broad-band spectral analysis of 2S 0921-63 has been done up to 25 keV. The 0.5--10 keV XIS count rate varied between $\sim$ 1 and $\sim$ 5 counts s$^{-1}$ during the observations. A partial X-ray eclipse and broad post-eclipse intensity dip were observed during the observations. The X-ray emission hardened marginally during the intensity dip. We have modelled the source spectra by simultaneously fitting the XIS and HXD-PIN spectra for each of the four observations. The broad-band spectra of the source can be described by a model comprising a very hot blackbody having temperature, $kT_{\rm BB} \approx 1.66-2.13$ keV, a high-energy cutoff power law, and an Fe emission line at $E_{\rm line} \sim 6.7$ keV. A second model, accounting for the Comptonization of the thermal emission from accretion disc along with an Fe emission line, describes the broad-band spectra of 2S 0921-63 equally well.    
\end{abstract}

\begin{keywords}
stars: neutron -- binaries: eclipsing -- X-rays: binaries -- X-rays: individual: 2S 0921-63\end{keywords}



\section{Introduction}

2S 0921-63 is an eclipsing low-mass X-ray binary (LMXB) discovered with \emph{SAS-3} \citep{Li1978}. The low-mass binary companion is an optical star of spectral type K0III \citep{Branduardi1983,Shahbaz1999}. From photometric and spectroscopic studies, \citet{Branduardi1981,Branduardi1983} and \citet{Cowley1982} determined an orbital period of about 9 d. Combined with \emph{EXOSAT} X-ray observations, \citet{Mason1987} refined the orbital period to 9.0115(5) d. Intensity dips and partial eclipses lasting for about 1.3 d have been observed in the optical \citep{Branduardi1983,Mason1985,Mason1987} as well as X-ray energies \citep{Mason1987}.  

The nature of the compact object is still unclear, mainly due to the absence of coherent pulsation or type-I X-ray bursts, typical to neutron star (NS) LMXBs. The initial mass estimate of 2.0--4.3 M$_{\sun}$ for the primary suggested the possibility of a massive NS or a low-mass black hole in the system \citep{Shahbaz2004,Jonker2005}. Later, \citet{Shahbaz2007} and \citet{Steeghs2007} refined the primary mass limits to lower values consistent with the canonical NS mass, thus supporting the idea of a NS primary.

2S 0921-63 exhibits an X-ray to optical flux ratio ($L_{X}/L_{opt}$) of the order 1, unusually less compared to $L_{X}/L_{opt} > 100$ for other sources \citep{Mason1985,Mason1987}. The possibility of an extended X-ray source obscured from direct view by an accretion disc at high inclination was proposed to explain the observed partial eclipses and low ratio of X-ray to optical flux \citep{Mason1985,Krzeminski1991}.


The 0.6--10 keV spectrum of 2S 0921-63 obtained with data from the \emph{Einstein} satellite was described with a thermal bremsstrahlung model having $kT \sim 13$ keV, moderately absorbed by a column density, $N_{\rm H} \sim 5.7 \times 10^{21}$ cm$^{-2}$ \citep{Branduardi1983}. The spectral analysis for the 1985 \emph{Einstein} data suggested the softening of spectrum during the eclipse \citep{Mason1987}. The spectrum showed similarity to the unsaturated Comptonized spectra, typical to other bright sources \citep[e.g., Sco X-1, Sco X-2, GX 17+2, X 1705-440;][]{White1985,White1986}. These results were inconclusive due to contamination of data at low energies.

The \emph{ASCA} spectrum showed significant hardening compared to the low-inclination LMXBs along with the presence of Fe emission line at 6.75 keV \citep{Asai2000}. \citet{Kallman2003} used the high-resolution data from \emph{Chandra} and \emph{XMM} and found the emission lines from O, Ne, Mg, Si, S, and Fe in the source spectrum.


In this paper, we present the broad-band spectral study of LMXB 2S 0921-63 by analysing the \emph{Suzaku} archival data obtained in August 2007. We aim to study the properties of the entire spectrum covering an energy range of 0.5--25 keV.

\begin{table*}
	\caption{Observation log of \emph{Suzaku} data.}
	\label{tab:obslog}
	\begin{tabular*}{\textwidth}{@{\extracolsep{\fill}} cccc|c|c|c|c|c}
	\hline
		Observation & Observation ID & Observation Date & MJD (d) & \multicolumn{2}{c|}{Exposure$^{a}$ (ks)} & \multicolumn{2}{c|}{Count Rate$^{b}$ (c/s)} & Orbital Phase$^{c}$\\
		& & dd-mm-yyyy &  & XIS & PIN & XIS & PIN &\\ 
		\hline
		1 & 402059010 & 23-08-2007 & 54335.89  & 43.13 & 34.74 & $3.023 \pm 0.008$ & $0.043 \pm 0.003 $ & 0.31--0.44\\[0.5ex]
		2 & 402060010 & 26-08-2007 & 54338.09  & 40.34 & 35.64 & $ 3.520 \pm 0.009 $ & $0.044 \pm 0.003 $ & 0.56--0.68\\[0.5ex]
		3 & 402057010 & 28-08-2007 & 54340.80  & 43.21 & 14.89 & $ 2.265 \pm 0.007 $ & $ 0.039 \pm 0.005 $ & 0.86--0.98\\[0.5ex]
		4 & 402058010 & 30-08-2007 & 54342.41  & 45.69 & 40.48 & $2.326 \pm 0.007 $ & $0.024 \pm 0.003 $ & 1.04--1.16 \\[0.5ex]
	\hline
	\multicolumn{8}{l}{\textit{Notes.} $^{a}$Exposure for the cleaned XIS and HXD-PIN spectra.}\\
    \multicolumn{8}{l}{$^{b}$Average count rate for XIS0, XIS1, and XIS3 for 0.5--10 keV energy range.}\\
    \multicolumn{8}{l}{$^{c}$Orbital phase based on the ephemeris of \citet{Ashcraft2012}.}\\
	\end{tabular*}
\end{table*}

\section{Observations}

The X-ray Imaging Spectrometer \citep[XIS;][]{Koyama2007} onboard the fifth Japanese X-ray observatory \emph{Suzaku} \citep{Mitsuda2007} comprises of four units numbered from 0 to 3. The three units are front-illuminated CCDs (XIS0, XIS2, and XIS3), and one is a back-illuminated CCD (XIS1). XIS provides an energy coverage in the range of 0.2--12 keV. The Hard X-ray Detector \citep[HXD-PIN;][]{Takahashi2007} extends the energy range to about 60 keV. \emph{Suzaku} observed 2S 0921-63 four times between starting 2007 August 23 till 2007 August 30 in normal modes and collected data in $3\times3$ and $5\times5$ pixel format (except for the second observation where only $3\times3$ mode data were present) for a combined exposure of 172.4 ks by using the XIS. HXD-PIN collected data for a net exposure of 125.8 ks. Table~\ref{tab:obslog} lists the details of the observations used in this work. We have utilized the data from XIS0, XIS1, and XIS3 detectors in our analysis. We do not use XIS2 data owing to damage due to micro-meteorites in 2006.

\subsection{XIS Spectra}
\label{sec:XIS Spectra} 
The unfiltered XIS event files have been screened and filtered utilizing the task \textsc{aepipeline} provided with the \textsc{heasoft} version 6.29a. We used the calibration database (CALDB) version 20181010 on processed XIS event files for the calibration. We used the \textsc{xselect} tool to extract the product files for each CCD individually, by combining the 3$\times$3 and 5$\times$5 pixel format cleaned event files. For each XIS detector, we used a circular region of 180 arcsec centred at the source to extract the source spectra. A similar-sized circular region was used to extract the background spectra situated away from the source photons. We generated the response files for each XIS detector by using the \textsc{xisrmfgen} tool. We used source coordinates R.A. $= 140.6444^\circ$ and Dec. $= 63.2948^\circ$ \citep{Cat2020} to produce the ancillary response files by using the script \textsc{xissimarfgen}. 

The spectra from all the detectors for each observation have been re-binned for at-least 20 counts per energy bin by using the ftool \textsc{grppha}. We used the spectra from all three XIS detectors for the simultaneous fitting in the energy range 0.5--10 keV of all four observations. We removed the energy bins between 1.7 and 2.3 keV to avoid the calibration uncertainties due to the artificial structures around Si and Au edges \citep{Jiang2021}.

\subsection{HXD-PIN Spectra}
 We have utilized the HXD-PIN data extending the energy coverage above 10 keV for the first time for the source. We extracted the HXD-PIN spectra for each observation following the standard procedure described in the \emph{Suzaku} ABC Guide\footnote{https://heasarc.gsfc.nasa.gov/docs/suzaku/analysis/abc/abc.html}. The raw event files were processed by using the HXD CALDB version 20110913. We generated the HXD-PIN spectra for each observation by using the \textsc{hxdpinxbpi} routine. The task uses pseudo-events files distributed with the archival data and performs dead time correction to the spectrum. We used observation-specific tuned PIN background event files to account for the contribution from the Non-X-ray Background and Cosmic X-ray Background (CXB). The \textsc{hxdpinxbpi} task simulates and models the CXB contribution and generates the total background spectrum following the model given by \citet{Boldt1987}. We used the response file \textit{ae\_hxd\_pinxinome4\_20080129.rsp} for HXD-PIN spectral fitting specific to the epoch of the observation. 
 
\begin{figure}
    \includegraphics[width=\columnwidth]{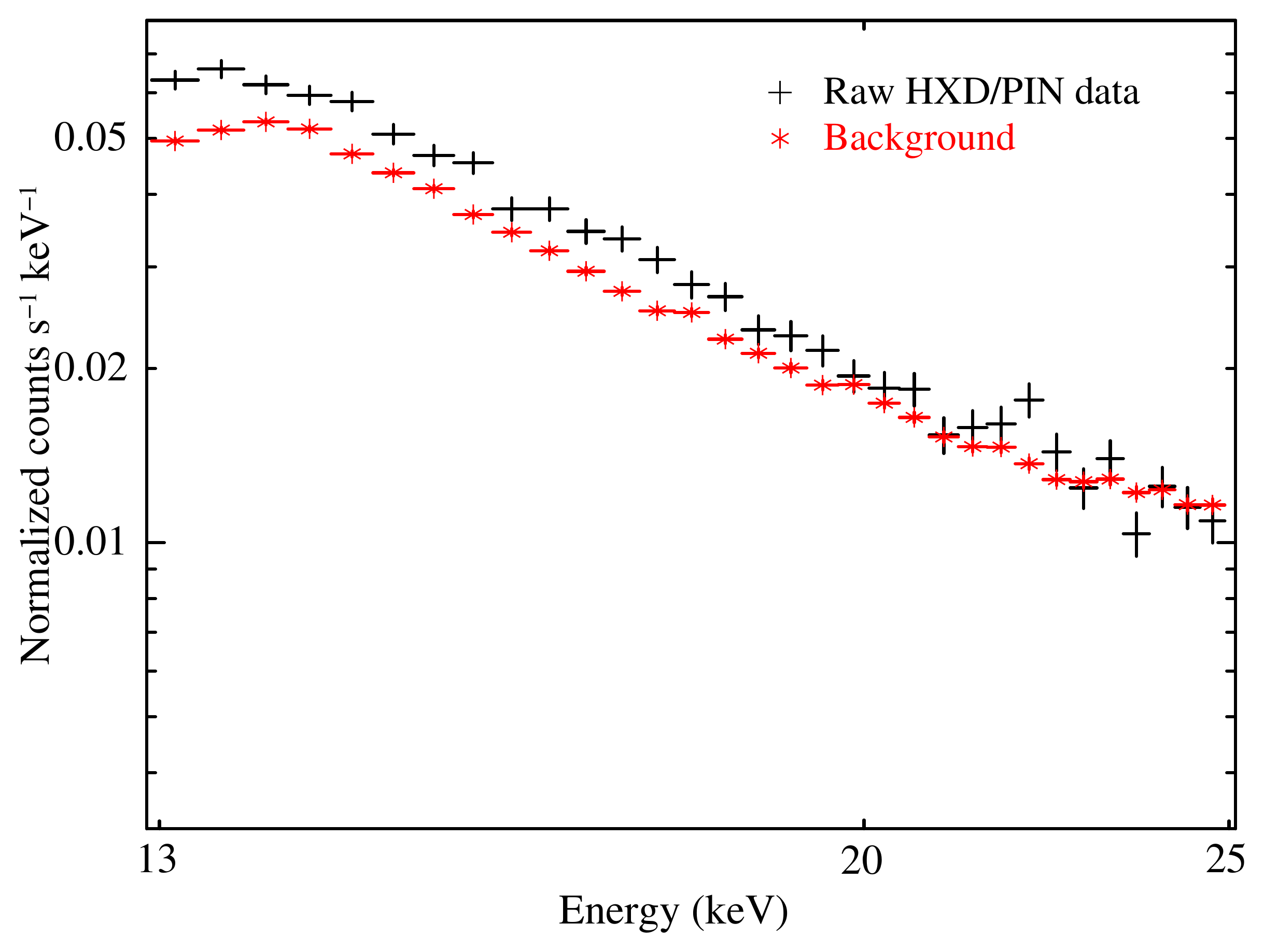}
    \caption{Raw source and background spectra of 2S 0921-63 observed with the HXD-PIN. The source count rate is significant for energies up to 25 keV.}
    \label{fig:back}
\end{figure}

 HXD-PIN detected the source up to 25 keV above the background during all four observations (Fig.~\ref{fig:back}). However, due to the low statistics below 13 keV, we ignored the energy bins below 13 keV and have used the 13--25 keV energy range for the spectral analysis.

\subsection{Light Curve}
\begin{figure}
    \includegraphics[width=\columnwidth]{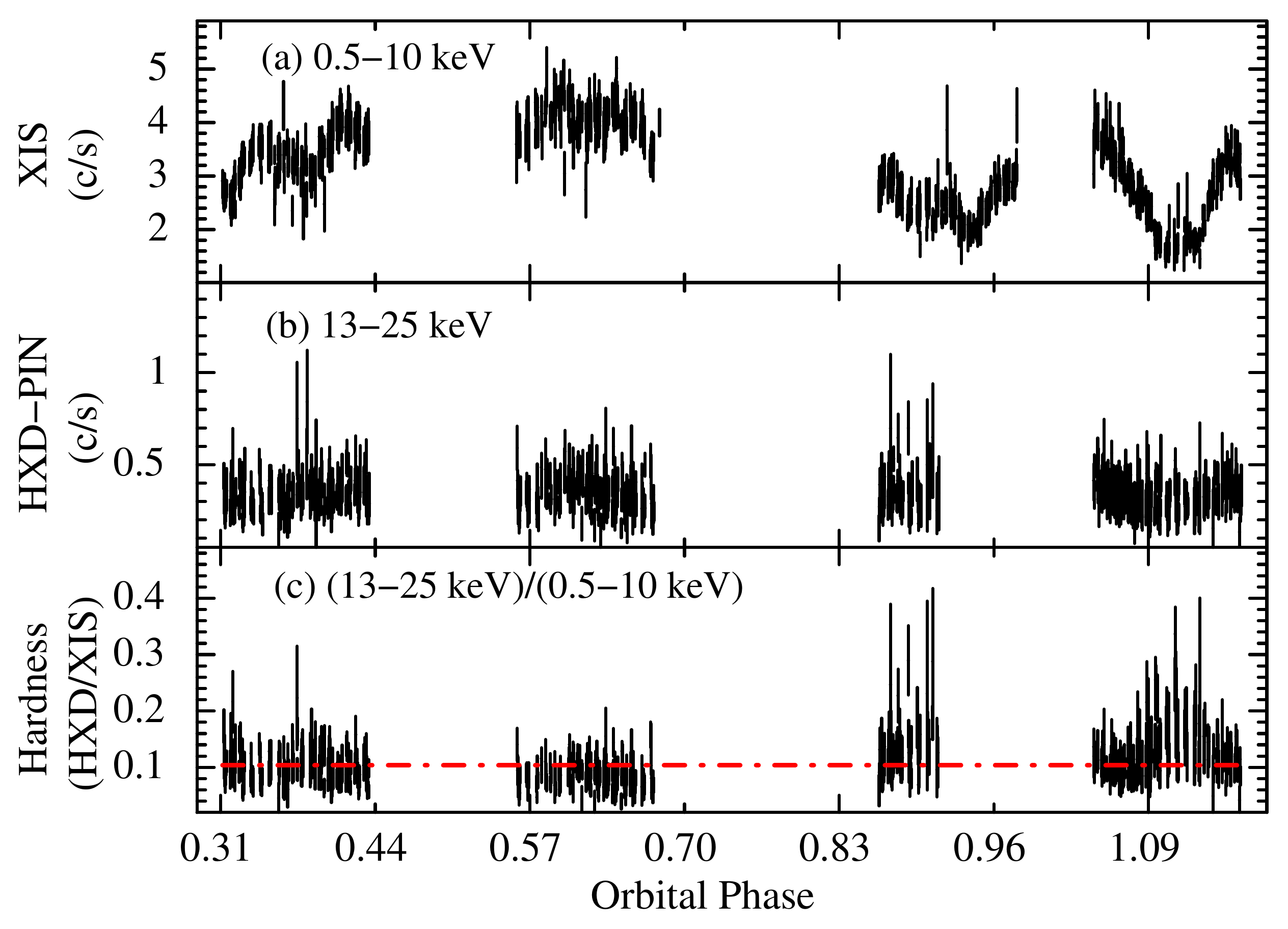}
    \caption{Background subtracted light curves of 2S 0921-63 binned at 128 s. The top panel shows the XIS light curve for 0.5--10 keV energy range. The middle panel gives the HXD-PIN light curve in 13--25 keV. The last panel shows the hardness ratio between HXD-PIN and XIS count rate.}
    \label{fig:light}
\end{figure} 
\begin{figure}
	\includegraphics[width=\columnwidth]{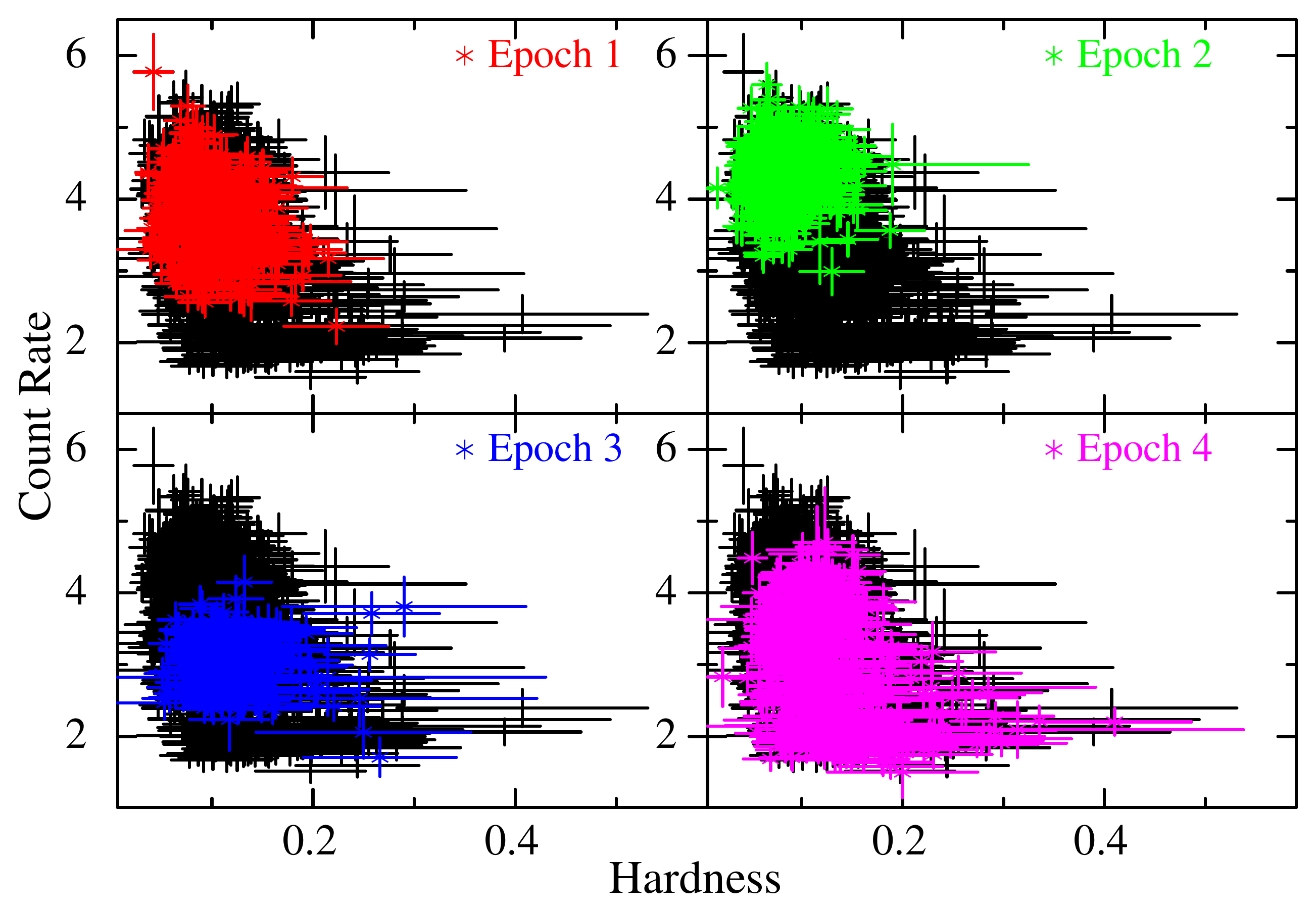}
	 \caption{Hardness-intensity plot of 2S 0921-63 of the four \emph{Suzaku} observation epochs with the hardness-intensity plot of the collective data in the background}. The XIS light curves in 0.5--10 keV and HXD-PIN light curves in 13--25 keV have been used for the soft and hard energy bands, respectively.
    \label{fig:hid}
    \end{figure} 
The XIS and HXD-PIN light curves in the respective energy bands of 0.5--10 keV and 13--25 keV are shown in Figure~\ref{fig:light}. The top and middle panels show the XIS and HXD-PIN light curves with corresponding hardness ratio in the bottom panel. The observations cover the orbital phases of 0.31--0.44, 0.56--0.68, 0.86--0.98, and 1.04--1.16, respectively, based on the ephemeris of \citet{Ashcraft2012} with phase zero at the centre of the eclipse. X-ray flux is variable across the four epochs. XIS count rate varies from $2.86 \pm 0.03$ counts s$^{-1}$ to $3.74 \pm 0.02$ counts s$^{-1}$ between the starting and ending of the first observation while the HXD-PIN count rate remains constant at $\sim 0.3$ counts s$^{-1}$.


During the second epoch, count rates increase relatively and remain considerably constant at $\sim$ 4 and $\sim$ 0.32 counts s$^{-1}$ for XIS and HXD-PIN, respectively. The orbital phase for the third epoch coincides with the eclipse of the primary X-ray source. Gradual decrease and a dip near the end of the observation are evident from the XIS light curve. The XIS count rate gradually decreases from $2.86 \pm 0.03$ to $2.40 \pm 0.03$ counts s$^{-1}$ and then to $2.00 \pm 0.04$ counts s$^{-1}$ near the dip. It then recovers to $2.92 \pm 0.05$ counts s$^{-1}$ near the end of the third observation. HXD-PIN collected data only for the first 15 ks and did not record any significant variation in the count rate near the dip. 

The last epoch shows significant variation with a broad dip in the XIS energy band consistent with the previously reported dips \citep{Mason1987,Krzeminski1991}. A marginal variation in the HXD-PIN count rate is also evident, resulting in the increased hardness near the centre of the dip. Similar hardening during the intensity dips has been reported for several eclipsing (e.g., X 2127+119: \citet{Ioannou2002}; XTE J1710-281: \citet{Younes2009}) and dipping sources (e.g., 4U 1323-62: \citet{Boirin2005}; X 1254-690: \citet{Diaz2009}). For a clearer picture, we generated a hardness intensity plot by using the light curves to compare the variation during different observation epochs (Figure~\ref{fig:hid}). While the source emission was variable during the span of the observations, though the variation in the hardness ratio was not significant. Thus, we used the time-averaged spectra from each observation for our analysis.

\section{Results}
\begin{figure}
   \includegraphics[width=\columnwidth]{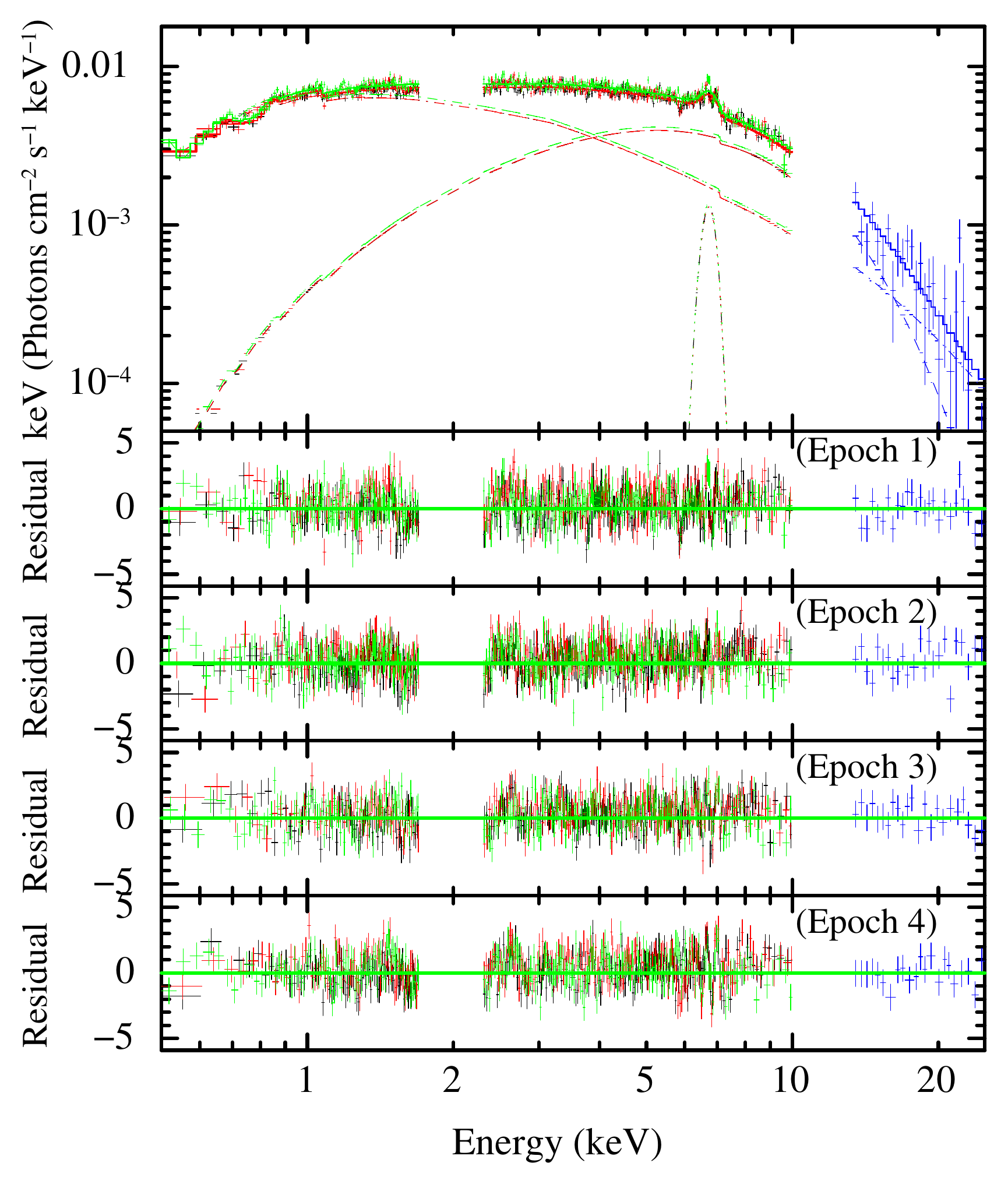}
\caption{The broad-band XIS (0.5--10 keV) and HXD-PIN (13--25 keV) spectra of 2S 0921-63. The upper panel gives the best-fitting unfolded spectrum modelled with M1 of the first observation. The lower panels show the residuals with respect to the model M1 of each observation. The black, red, and green markers represent XIS0, XIS3, and XIS1 spectra, respectively. HXD-PIN spectra are marked with blue data points.}
 \label{fig:spec1}
\end{figure}

We have used \textsc{xspec} \citep{Arnaud1996} v12.12.0 for simultaneous spectral fitting of XIS and HXD-PIN spectra for each observation individually. We introduced a cross-normalization factor to account for the different instrumental calibrations. We fixed its value at 1 for the XIS0 and XIS3 data sets and left it free for XIS1. The cross-normalization factor was fixed at 1.158 for HXD-PIN \citep{Kokubun2007}. We have adopted the updated photo-ionization cross-sections of \citet{Verner1996} and solar abundances by \citet{Wilms2000} for our spectral analysis. We tied all the model parameters across XIS0, XIS3, XIS1, and HXD-PIN spectra for spectral fitting. 

The X-ray spectrum of 2S 0921-63 has been studied only up to 10 keV. The spectrum is generally modelled with an absorbed simple power law or cutoff power law along with emission features from different elements \citep{Mason1987,Kallman2003}. Following this, we tried to fit the spectra by using the high-energy cutoff power law model. We included the \texttt{tbabs} component to account for the absorption medium. This model failed to provide an adequate fitting to the spectra with large systematic residuals below 1.5 keV and between 6--7 keV for all four epochs. The addition of a Gaussian component (\texttt{gaus}) at 6.7 keV gave a statistically better fit. But a large residual below 1.5 keV still existed along with a narrow feature near 7 keV. 

We then used a two-component model, typical for LMXBs, i.e., a soft thermal component with a power law component. We added the \texttt{bbody} \citep{Mitsuda1984} component to the existing model to account for any contribution from the NS. It improved the fittings significantly for all the four epochs. The best-fitting returned blackbody temperature between 1.7 and 1.8 keV, photon index in the range 1.6--1.7, cutoff energy ($E_{\rm cut}$) $\sim 3$ keV, and Fe emission line energy $\sim 6.7$ keV. Although the new model provided a reasonable good fitting to the spectra, some features could not be modelled. Some residual persisted around 0.7 keV, 1 keV, and 7 keV. While we identified the latter ones as possible absorption edges from Ne (1.196 keV) and Fe (7.117 keV), we used a Gaussian absorption component to model the 0.7 keV feature in the spectra from all epochs except the third epoch, where it was not statistically favoured. Contrary to our claim of an absorption edge at 1.196 keV, \citet{Kallman2003} identified the 1.17 keV feature in \emph{Chandra} and \emph{XMM} data as the Ne \textsc{x} Ly $\beta$ emission line. But due to the limited resolution of \emph{Suzaku} data and model uncertainties, it is not possible to determine the actual nature of this feature.


\begin{figure}
   \includegraphics[width=\columnwidth]{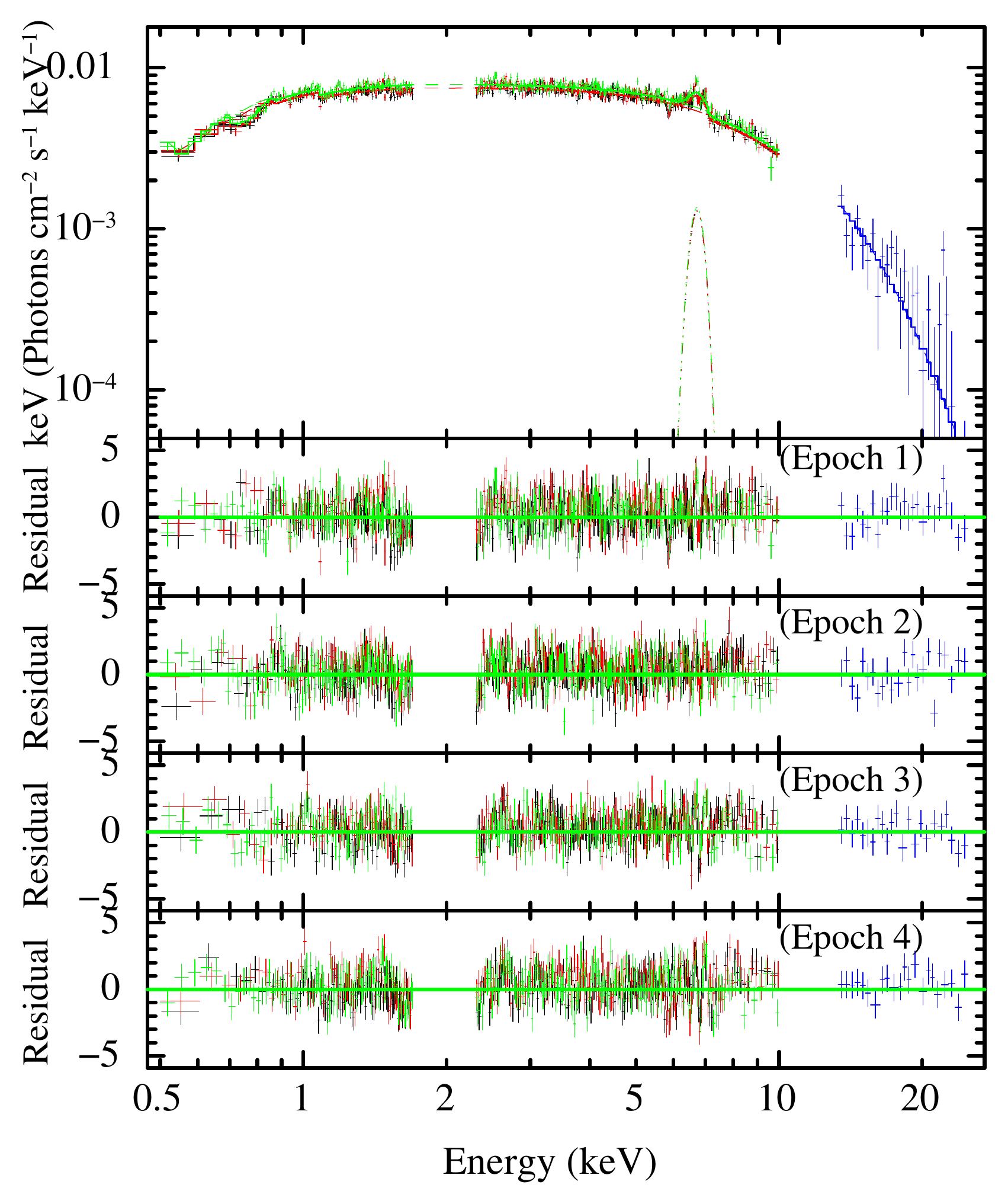}
\caption{The broad-band XIS (0.5--10 keV) and HXD-PIN (13--25 keV) spectra of 2S 0921-63. The upper panel gives the best-fitting unfolded spectrum modelled with the Comptonization model, M2 of the first observation. The lower panels show the residuals with respect to the model M2 of each observation.}
 \label{fig:spec2}
\end{figure}

\begin{table*}
	\caption{Best-fitting spectral parameters for broad-band \emph{Suzaku} spectra of 2S 0921-63 of four observations. The errors are quoted at 90 per cent confidence level.}
	\label{tab:spec}
\begin{tabular}{c | cccc | cccc}
    \hline
    &\multicolumn{4}{c}{Model M1} & \multicolumn{4}{c}{Model M2} \\
    \hline
    Parameters & Epoch 1 & Epoch 2 & Epoch 3 & Epoch 4 & Epoch 1 & Epoch 2 & Epoch 3 & Epoch 4 \\
    \hline
    $C_{\rm XIS1}$ & $1.05 \pm 0.01$ & $1.04 \pm 0.01$ & $1.04 \pm 0.01$ & $1.04 \pm 0.01$ &  $1.05 \pm 0.01$ & $1.04 \pm 0.01$ & $1.04 \pm 0.01$ & $1.04 \pm 0.01$ \\[1ex]
     
   $N_{\rm H}$ ($10^{22}$ cm$^{-2}$) & $0.27 \pm 0.02$ &  $ 0.30 \pm 0.02$ & $0.15_{-0.06}^{+0.03}$ & $0.29 \pm 0.02$ & $0.21_{-0.02}^{+0.01}$ & $0.20_{-0.09}^{+0.02}$ & $0.22_{-0.03}^{+0.04}$  & $0.25_{-0.06}^{+0.02}$  \\[1ex]

    $E_{\rm Ne \ \textsc{ix} }$ (keV) & $1.07 \pm 0.01$ &$1.08 \pm 0.02$ & $1.08_{-0.03}^{+0.04}$ & $1.16_{-0.04}^{+0.03}$ & $1.08 \pm 0.02$ &$1.09 \pm 0.02$  & $1.10_{-0.03}^{+0.05}$ &  $1.17 \pm 0.02 $\\[0.5ex]
    $\tau$ & $0.10 \pm 0.02 $ &$0.12 \pm 0.02$ & $0.07 \pm 0.03$ & $0.07 \pm 0.02$ & $0.09_{-0.01}^{+0.02}$ & $0.11 \pm 0.02$ & $0.05 \pm 0.02$ & $0.09 \pm 0.02$ \\[1ex]
    
    $E_{\rm Fe \ \textsc{i} }$ (keV) & $7.08_{-0.07}^{+0.09}$ & $7.10 \pm 0.04$& $7.09 \pm 0.04$ & - & $7.09 \pm 0.07 $ & $7.10 \pm 0.05$ & $7.09 \pm 0.04$ & - \\[0.5ex]
    $\tau$ & $0.07_{-0.03}^{+0.04}$ &$ 0.14 \pm 0.03$ & $0.17_{-0.03}^{+0.04}$ & - & $ 0.07 \pm 0.02$ & $0.11_{-0.01}^{+0.03}$ & $0.16 \pm 0.03 $ & - \\[2ex]
    
    $kT_{\rm BB}$ (keV) & $1.85_{-0.06}^{+0.08}$ & $1.94 \pm 0.04 $ & $2.05_{-0.07}^{+0.08} $ & $ 1.72 \pm 0.06$ & - & - & - & -\\[0.5ex]
    $N_{\rm BB} $ ($10^{-4}$) & $ 6.45_{-0.39}^{+0.67}$ & $8.34_{-1.16}^{+0.84}$ & $6.41_{-0.73}^{+0.53}$ & $3.80_{-0.37}^{+0.50}$ & - & - & - & - \\[0.5ex]
    $f^{\dag}_{\rm bol}$  & $5.40 \pm 0.06 $ & $6.96 \pm 0.07$ & $5.37 \pm 0.07$ & $3.19 \pm 0.04$ & - & - & - & - \\[2ex]
     
    POWERLAW  ($\Gamma$) & $1.71_{-0.05}^{+0.06}$ & $ 1.86 \pm 0.10 $ & $ 0.75 \pm 0.33$ & $1.66 \pm 0.06$  & - & - & - & -\\[0.5ex]
    Norm ($10^{-2}$) & $1.03 \pm 0.04 $ & $1.42 \pm 0.06 $ & $0.58 \pm 0.06$& $0.84 \pm 0.03$ &- & - & - & - \\[0.5ex]
    $E_{\rm cut}$ (keV) & $3.17_{-0.59}^{+0.27}$ & $ 3.28_{-0.14}^{+0.11}$ & $1.04_{-0.06}^{+0.07}$ & $3.08_{-0.49}^{+0.16}$ & - & - & - & - \\[0.5ex]
    $E_{\rm f}$ (keV) & $8.33_{-2.19}^{+2.17}$ & $4.24_{-1.16}^{+1.99}$ &$2.22_{-0.70}^{+1.44}$ & $7.17_{-1.57}^{+1.65}$ & - & - & - & -  \\[0.5ex]
    $f^{\dag}_{\rm bol}$ &$8.16 \pm 0.71$ & $9.45 \pm 0.05 $ & $3.42 \pm 0.06$ & $6.53 \pm 0.04$ & - & - & - & -  \\[2ex]
    
    NTHCOMP  ($\Gamma$) & - & - & - & - & $1.46 \pm 0.01$ & $1.52 \pm 0.01$ & $1.46 \pm 0.01$ & $1.51 \pm 0.01 $ \\[0.5ex]
    $kT_{\rm e}$ (keV) & - & - & - & - & $2.12 \pm 0.04$ & $2.14 \pm 0.04$ & $2.25_{-0.04}^{+0.07}$ & $1.94_{-0.03}^{+0.04}$ \\[0.5ex]
    $kT_{\rm seed}$ (keV) & - & - & - & - & $0.20 \pm 0.06$ & $0.20_{-0.04}^{+0.03}$ & $ < 0.39 $ & $ < 0.41 $ \\[0.5ex]
    $R_{\rm o}$ (km) & - & - & - & - & $45_{-20}^{+50}$ & $ 50_{-14}^{+31}$ & $ > 10$ & $> 10 $ \\[0.5ex]
    $f^{\ddag}_{\rm bol}$ &  &  &  &  & $1.20 \pm 0.01$ & $1.38 \pm 0.01 $ & $0.93 \pm 0.01$ & $0.87 \pm 0.01$  \\[2ex]

    $E_{\rm Fe \textsc{xxv}}$ (keV) & $ 6.72 \pm 0.03$ & $ 6.69 \pm 0.03 $ & $6.78_{-0.03}^{+0.04}$ & $6.74 \pm 0.02$ & $ 6.71 \pm 0.03$ & $ 6.69 \pm 0.03$ & $ 6.78 \pm 0.03$ & $ 6.74 \pm 0.02$\\[0.5ex]
    $\sigma$ (keV) & $0.23 \pm 0.03$ & $ 0.23_{-0.03}^{+0.04} $ & $ 0.21 \pm 0.03$ & $ 0.21 \pm 0.02$ & $0.23 \pm 0.03$ & $ 0.24 \pm 0.03 $ & $ 0.22_{-0.03}^{+0.04} $ & $ 0.20 \pm 0.02 $ \\[0.5ex]
    Norm ($10^{-4}$) & $ 1.15_{-0.16}^{+0.14}$ & $1.13 \pm 0.17 $ & $1.01_{-0.12}^{+0.14}$ &  $1.04 \pm 0.10 $&  $1.13 \pm 0.14 $ & $1.22 \pm 0.15$ & $0.96 \pm 0.13$ & $1.02 \pm 0.10 $ \\[0.5ex]
    EW (eV) & $142_{-25}^{+29}$ & $128_{-26}^{+24}$ & $147_{-20}^{+35}$ & $195_{-19}^{+22}$ & $< 160 $ & $140_{-24}^{+23}$ & $156_{-40}^{+29} $ & $ < 207 $ \\[2ex]

    $E_{\rm abs}$ (keV) & $ 0.77 \pm 0.02$ & $ 0.76_{-0.01}^{0.02} $ & - & $0.78 \pm 0.02$ & $ 0.76 \pm 0.02$ & $ 0.73_{-0.04}^{+0.03}$ & - & $ 0.78 \pm 0.02$\\[0.5ex]
    $\sigma$ (keV) & $0.04 \pm 0.02$ & $ 0.05 \pm 0.02 $ & - & $0.04 \pm 0.02$ & $0.04_{-0.01}^{+0.02}$ & $ 0.06 \pm 0.02 $ & - & $0.04 \pm 0.02$ \\[0.5ex]
    Norm ($10^{-5}$) & $ 2.9_{-0.9}^{+1.0}$ & $4.48_{-1.4}^{+1.8} $ & - & $22.6_{-7.6}^{+9.5}$ & $2.4_{0.6}^{+0.9}$ & $ 46.4_{-30.3}^{+16.1}$ & - & $1.8_{-0.8}^{+0.6}$\\[0.5ex]
    EW (eV) & -$(17_{-4}^{+3})$ & -$(20_{-4}^{+3})$  & - &-$(17_{-5}^{+4})$ & -$(< 21)$ & -$(25_{-5}^{+4})$ & - & -$(< 20)$ \\[2ex]

    $f^{\ddag}_{\rm Total}$ &$ 1.36 \pm 0.01 $ &$1.65 \pm 0.04$ & $0.89 \pm 0.03$& $0.98 \pm 0.03$ & $1.21 \pm 0.01$ & $1.38 \pm 0.01$ & $0.94 \pm 0.01$ & $0.88 \pm 0.01$ \\[1ex]
    $L^{\S}_{\rm X}$ &$ 0.79 \pm 0.01 $ &$ 0.97 \pm 0.02$ & $0.52 \pm 0.02$& $0.57 \pm 0.02$ & $0.71 \pm 0.01$ & $0.81 \pm 0.01$ & $0.56 \pm 0.01$ & $0.51 \pm 0.01$ \\[1ex]
    
    \hline
    $\chi^2$/d.o.f & $5986.3/5643$ & $5694.7/5606$ & $5317.1/5367$ & $5473.5/5304$ & $5998.8/5645$ & $5730.9/5608$ & $5327.7/5369$ & $5491.1/5306$ \\
    \hline
	\multicolumn{9}{l}{\textit{Note.} Model M1 = \texttt{tbabs*edge*edge*(bbody +powerlaw*highecut + gaus + gaus)}} \\
    \multicolumn{9}{l}{Model M2 = \texttt{tbabs*edge*edge*(nthcomp[diskbb] + gaus + gaus)}} \\
	\multicolumn{9}{l}{$^{\dag} f_{\rm bol}$ is the unabsorbed flux in the energy band 0.1--100 keV in units of $10^{-11}$ erg cm$^{-2}$ s$^{-1}$ .}\\
	\multicolumn{9}{l}{$^{\ddag} f_{\rm Total}$ is the unabsorbed flux in the energy band 0.1--100.0 keV in units of  $10^{-10}$ erg cm$^{-2}$ s$^{-1}$.}\\
	\multicolumn{9}{l}{$^{\S} L_{\rm X}$ is the unabsorbed luminosity in the energy band 0.1--100.0 keV in units of  $10^{36}$ erg s$^{-1}$.}\\
	\end{tabular}
\end{table*}

We, thus, define \texttt{tbabs*egde*edge*(bbody + powerlaw*highecut + gaus + gaus)} in \textsc{xspec} as our first model, M1. The model provided an acceptable fitting to the spectra with a $\chi^2$/$\nu$ of 5986.3/5643, 5694.7/5606, 5317.1/5367, and 5473.5/5304 for the four observations, respectively. Figure~\ref{fig:spec1} shows the best-fitting unfolded spectra for the first epoch with model M1 and its components. The lower panels show the residuals with respect to model M1 of the four observations. All the best-fitting parameters are reported in Table~\ref{tab:spec}. 

Our attempts to model the spectra using a cutoff power law instead of \texttt{highecut} power law failed as the fitting resulted in substantial residuals near the lower and higher energy end of the spectra. When we replaced \texttt{bbody} component with the multi-colour accretion disc blackbody, \texttt{diskbb}, the fitting returned unreasonably high disc temperatures $\sim$ 3 keV with very low disc normalization values thus, we did not include the results for these models in our analysis. 

The typical two-component model successfully modelled the broad-band \emph{Suzaku} spectra of 2S 0921-63. From literature, it is known that the spectrum of 2S 0921-63 can be described as an unsaturated Comptonized spectrum, typical to X-ray sources (e.g., Sco X-1:  \citep{Mason1987}; Sco X-2, GX 17+2: \citet{Krzeminski1991}). 

Utilizing the broad-band coverage of \emph{Suzaku}, we check for the Comptonized emission from the source. We fit the spectra by using the thermal Comptonization model \texttt{nthcomp} \citep{Zdziarski1996, Zycki1999} in \textsc{xspec}. We included two Gaussian components and two edges to the model as before and set the seed photon source to NS blackbody. The model provided a reasonable fitting to the spectra however the fits returned low values of $kT_{\rm seed}$ ($\sim$ 0.2 keV) and a large emission radius ($>$ 40 km). We then changed the seed photon source to the accretion disc. This model, \texttt{tbabs*edge*edge*(nthcomp[dbb] + gaus + gaus)} (M2, hereafter) provided an acceptable fitting to the spectra of all four observations. The best-fitting spectra and residuals are shown in Figure~\ref{fig:spec2}, with derived parameters listed in Table~\ref{tab:spec}. We have computed the unabsorbed bolometric flux in 0.1--100 keV for the model and components by using the convolving model \texttt{cflux}. 


\section{Discussion}

We have analysed the data of LMXB 2S 0921-63 obtained with \emph{Suzaku} in August 2007. We have performed a broad-band spectral analysis of the source by using about 172 ks of data. We have used the XIS data covering 0.5--10 keV for this purpose. We have also, for the first time, extended the energy coverage for the spectral analysis to 25 keV by using the HXD-PIN (13-25 keV) data. We have modelled the spectra with a two-component model and have used a Comptonization model to describe the spectra. 

We have presented the XIS (0.5--10 keV) and HXD-PIN (13--25 keV) light curves covering the orbital phases between 0.31 and 1.16. The light curve covers the partial eclipse (0.86--0.98) and includes the post-eclipse dipping phase. While the eclipse is not completely covered by HXD-PIN, it is evident from the hardness ratio plot that the emission hardens marginally during the dip.The smooth shape of the dip during the last observation and its proximity to the predicted eclipse makes it difficult to establish if the intensity dip during the third observation is the original eclipse. Similarly, \citet{Krzeminski1991} reported the presence of many identical eclipses in the light curve of 2S 0921-63 during one cycle. Until a new X-ray orbital ephemeris is established, it will be difficult to identify the real partial eclipse confidently.

We have modelled the broad-band spectra of 2S 0921-63 by using a single temperature blackbody to check for the presence of NS blackbody emission \citep{Lyu2014,Zhang2016} and a high-energy cutoff power law component to account for the non-thermal contribution. For all four observations, we obtained a high blackbody temperature of $1.85_{-0.06}^{+0.08}$, $1.94 \pm 0.04 $, $2.05_{-0.07}^{+0.08} $, and $ 1.72 \pm 0.06$ keV, respectively. We interpret this hot blackbody component as the NS surface/boundary layer between disc \citep{Popham2001,Cackett2010,Sharma2020}. The best-fitting photon indices decrease, suggesting that the spectra harden from the first observation towards the last observation (Table~\ref{tab:spec}). The cutoff energy ($E_{\rm cut}$) attains a low and constant value $\sim 3$ keV during all observations except for the third observation, where it falls significantly to $1.04_{-0.06}^{+0.07}$ keV. The low value of $E_{\rm cut}$ suggests a low corona temperature \citep{Zhang2016}. The best-fitting confirmed a low neutral column density with $0.27 \pm 0.02$, $ 0.30 \pm 0.02$, $0.15_{-0.06}^{+0.03}$, and $0.29 \pm 0.02 \times 10^{22}$ cm$^{-2}$ for four observations, respectively. These values are in agreement with the previously reported values of column density \citep{Mason1987,Kallman2003}.

\begin{figure*}
   \includegraphics[width=1.8\columnwidth,height=0.7\columnwidth]{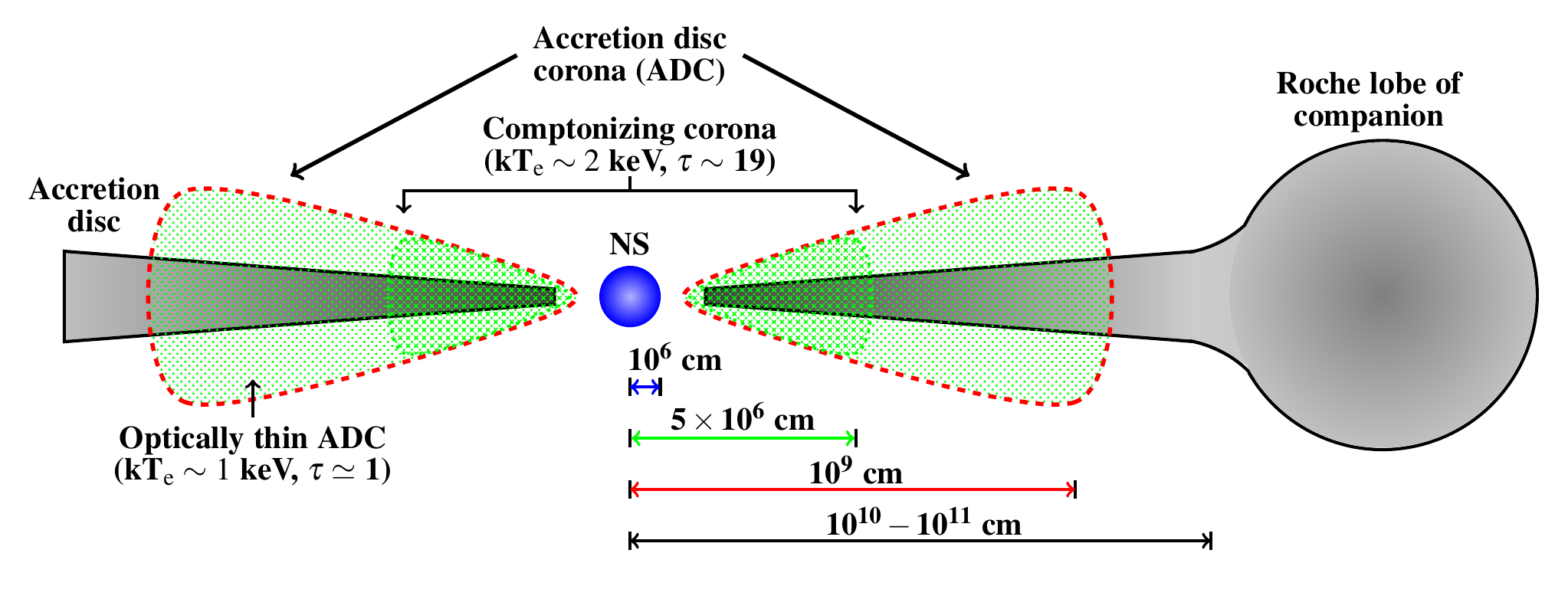}
\caption{The schematic diagram showing the cross-sectional view of the ADC source 2S 0921-63. The densely dotted inner region represents an optically thick ADC.}
 \label{fig:ADC}
\end{figure*}

The spectra of LMXBs are often described with the Comptonization model as it provides more information than a simple power law description \citep{Iaria2005,Sakurai2012,Zhang2014,Zhang2016,Sharma2020}. We used the broad-band coverage of \emph{Suzaku} and modelled the entire spectrum of 2S 0921-63 with a Comptonization model for the first time. A single component disc-seed Comptonization model provided a satisfactory good fitting to the spectra of all four observations. The photon index values agreed with the model M1 results favouring a hard spectral state. A low corona temperature, $kT_{\rm e} \sim 2$ keV, is again consistent with the low $E_{\rm cut}$ value from model M1. We obtained a low seed photon temperature of $0.20 \pm 0.06$ and $0.20 \pm 0.03$ keV for the first two observations with an upper limit of 0.39 and 0.41 keV for the last two observations. For the Comptonizing medium, we have estimated a relatively high optical depth of $\sim 19$ \citep{Zdziarski1996}. This high value is possibly due to the Comptonized component being close to the Wien spectrum \citep{Cackett2010}. Similar low corona temperature of the order $\sim$ 2--3 keV with high optical depth (11--13) have been reported for LMXB 4U 1758-20 \citep{White1986}.

The size and location of the Comptonized component is not known convincingly. Some theories claim an inner compact region ($ \leq $ 100 km) of the binary \citep{Kluzniak1991,Popham2001} while others support a large extended region ( $\approx$ 50000 km) \citep{White1982,Church2001,Church2004}.

For a clearer picture of the ADC and Comptonization region, we estimated the size for the seed photon emission region by using the equation from \citet{intzand1999}. From the first two observations, we obtained a seed photon region of $\approx $ 45 and 50 km and a lower limit of 10 km from the last two observations. Clearly, the size of the Compton seed region is consistent with the source of seed photons being the inner disc region.

Moreover, a high value of optical depth ($\sim$ 19) implies that the Comptonized component originates from an optically thick corona extending over the accretion disc regions at least up to $\sim$ 50 km. A similar corona geometry has been reported for the ADC sources X 1624-490 \citep{Iaria2007} and 4U 1822-371 \citep{Anitra2021}. For high inclination systems, a flat Comptonizing corona geometry can explain the large optical depths as up-scattered photons travel longer path along the line of sight \citep{Zhang2014,Zhang2016}.

Following the radius-luminosity relation for ADC, $r_{\rm ADC}\ = \ L_{\rm X}^{0.88} $ by \citet{Church2004}, we estimate $r_{\rm ADC} \approx 10^9$ cm. Again, using their equation (3) for radius $r_{\rm c}$, for corona with temperature $kT_{\rm e}$ to remain in hydrostatic equilibrium gives, $r_{\rm c} \sim 10^9$ cm. Thus, both the estimates agree with ADC being extended and covering the disc up to distance of $\sim 10^9$ cm. However, the Comptonization component that we observe in the spectra dominantly originates from the inner, optically thick region of ADC. Based on these calculations, Figure~\ref{fig:ADC} shows a schematic diagram for a clearer picture of this ADC source.


The presence of broad Fe emission lines is reported often in LMXBs \citep{Bhattacharyya2007,Cackett2010,Miller2013,Degenaar2014}. We found a broad emission line around 6.7 keV in the spectra of all four observations. We used a Gaussian component to model this feature with both the models M1 and M2. The best-fitting line energy for the feature is consistent with the highly ionized Fe \textsc{xxv} (He-like) emission line. The origin of Fe emission lines in LMXBs is generally attributed to the recombination in the ionized matter around the NS \citep{White1985,White1986,Hirano1987,Kitamoto1987}. \citet{Kallman2003} also reported the presence of Fe \textsc{xxv} emission line along with neutral Fe and Fe \textsc{xxvi} lines in the spectra of 2S 0921-63. We have estimated upper limits of 11, 17, 13, and 11 eV for the neutral Fe line and 48, 36, 69, and 76 eV for Fe \textsc{xxvi} line on the equivalent widths, respectively for the four observations. While the equivalent width of the Fe \textsc{xxv} emission line is high compared to the previously reported value, the value is consistent across the four observations. 


We observed softening of the emission during the third observation covering the eclipse phase. The blackbody temperature increased as compared to the first two observations and reached about 2 keV. While the photon index and cutoff energy decrease, the thermal blackbody emission dominates with about $\sim 60$ per cent contribution to the total emission flux. During the last observation, the blackbody temperature decreases to $\sim 1.72$ keV with a hard photon index of $\sim 1.72$ and cutoff energy $\sim 3.8$ keV. The non-thermal component dominates with a contribution of $\sim 67$ per cent to the total flux. Typical of the variation in the light curve, the unabsorbed bolometric flux in 0.1--100.0 keV was maximum during the second observation and minimum during the highly variable last observation. Net flux during the third and fourth observations show a significant decrease as compared to the first two epochs, consistent with the eclipse and dip.

We have found an absorption-like feature around 1 keV in the spectra from all observations. We used an absorbed Gaussian component to model this feature around 0.7--0.8 keV in all the spectra except for the third observation, where it was not statistically favoured. Similar dips near 1 keV have been reported for other dipping sources (e.g., XB 1916-053, EXO 0748-676, X 1254-690, MXB 1658-298: \citet{Diaz2006}; XTE J1710-281: \citet{Younes2009}). It is possible that the feature arises from the blending of several lines and edges produced by the ionized absorber around the source. It could also be due to the known contamination of the XIS detectors, primarily because of the Oxygen below 2 keV. A similar feature was found in the \emph{Suzaku} spectra of XTE J1710-281 around 0.6 keV \citep{Sharma2020}. We also found an absorption edge at 7.1 keV, implying no or very low ionization of the absorbing matter \citep{Singh1994}. Such features may arise from the reflection of hard X-rays off the cold and optically thick material from the outer regions of the accretion disc \citep{Gondoin2001}.

Although the partial covering models did not provide an acceptable fitting to the current \emph{Suzaku} data, the presence of intensity dips does not rule out the possibility of an additional medium around the primary source. Future targeted observation of 2S 0921-63 near dipping phases may prove useful in understanding its properties better.    

\section*{Acknowledgements}

We have used the data obtained with \emph{Suzaku}, a JAXA/ISAS space mission, and used software provided by the High Energy Astrophysics Science Archive Research Center (HEASARC) online service maintained by the NASA Goddard Space Flight Center. We thank the anonymous referee for the useful suggestions that have helped in improving the results. PS acknowledges the financial support from the Council of Scientific \& Industrial Research (CSIR) under the Senior Research Fellowship (SRF) scheme. 

\section*{Data Availability}
The underlying work has used the data obtained with \emph{Suzaku} and is available at the High Energy Astrophysics Science Archive Research Center (HEASARC) online service. The archival data can be accessed at \url{https://heasarc.gsfc.nasa.gov/cgi-bin/W3Browse/w3browse.pl}.



\bibliographystyle{mnras}
\bibliography{2S0921} 





\bsp	
\label{lastpage}
\end{document}